\documentclass[aps,prd,reprint,preprintnumbers,longbibliography,nofootinbib]{revtex4-2}
\pdfoutput=1

\usepackage[hidelinks,colorlinks=true,linkcolor=blue,citecolor=blue]{hyperref}
\usepackage[utf8]{inputenc}
\usepackage{uniinput}
\usepackage{amsmath,amsfonts}
\usepackage{url}
\usepackage{bm}
\usepackage{mathtools}
\usepackage{relsize}
\usepackage{cancel}
\usepackage{lipsum,mathrsfs}
\usepackage{hyperref}
\usepackage[plain]{fancyref}
\usepackage{xcolor}
\usepackage{transparent}
\usepackage{enumitem}

\definecolor{newred}{HTML}{DC3220}

\renewcommand{\d}{\mathrm{d}}

\newcommand{\bea}{\begin{eqnarray}}
\newcommand{\eea}{\end{eqnarray}}
\newcommand{\be}{\begin{equation}}
\newcommand{\ee}{\end{equation}}

\usepackage{microtype}
\allowdisplaybreaks

\begin{document}
\title{Features of a dark energy model in string theory }

\author{Souvik Banerjee,}
\email{souvik.banerjee@physik.uni-wuerzburg.de}
\affiliation{Institut für Theoretische Physik und Astrophysik, Julius-Maximilians-Universität Würzburg, Am Hubland, 97074 Würzburg, Germany}

\author{Ulf Danielsson,}
\email{ulf.danielsson@physics.uu.se}
\affiliation{Institutionen för fysik och astronomi,
	Uppsala Universitet, Box 803, SE-751 08 Uppsala, Sweden}

\author{Suvendu Giri~}
\email{suvendu.giri@unimib.it}
\affiliation{Dipartimento di Fisica, Università di Milano-Bicocca, I-20126 Milano, Italy}
\affiliation{INFN, sezione di Milano-Bicocca, I-20126 Milano, Italy}

\begin{abstract}
\noindent
In this article we clear up misconceptions concerning the dark bubble model as a realization of dark energy in string theory. In particular we point out important differences with Randall-Sundrum, and explain why gravity neither is, nor need to be, localized on the dark bubble.
\end{abstract}

\preprint{UUITP-63/22}
\maketitle
\section{Introduction and overview}\label{sec:intro}

The dark bubble model \cite{Banerjee:2018qey,Banerjee:2019fzz,Banerjee:2020wix,Banerjee:2020wov,Banerjee:2021qei} (see \cite{Banerjee:2021yrb} for a quick review) was introduced as an alternative way to get dark energy from string theory, circumventing problems encountered by other attempts and its conjectured non-compatibility with quantum gravity \cite{Vafa:2005ui,Palti:2019pca,Grana:2021zvf,vanBeest:2021lhn} (see \cite{Danielsson:2018ztv} for an overview).\footnote{A realization of dS on a dark bubble in non-supersymmetric string theories was studied in \cite{Basile:2020mpt}. See \cite{Blaback:2019zig,Cordova:2018dbb,Cordova:2019cvf,DeLuca:2021pej,Berglund:2021xlm} for some other attempts at constructing dS in string/M-theory, and \cite{Berglund:2022qsb} for a recent overview.}
A crucial component, is the nucleation of bubbles of true vacuum in an initial unstable AdS$_5$ space time.\footnote{See \cite{Koga:2019yzj,Koga:2022opd} for studies of bubble nucleation catalysis in the presence of black holes or a cloud of strings.}
As shown in, \cite{Danielsson:2021tyb}, the nucleation event can be identified with the creation of the universe in Vilenkin's version of quantum cosmology. In \cite{Danielsson:2022lsl}, an explicit embedding of the model into string theory was found---including a natural explanation for the smallness of the cosmology constant.

Questions have been raised about whether the model consistently describes Einstein gravity in the appropriate limit. In this paper we will address these and other concerns and reiterate the importance of semi-infinite stretched strings in the bulk for obtaining spin-2 (Einstein) gravity in four dimensions. The paper is organized as follows. In \fref{sec:differences} we highlight some of the major differences between the dark bubble construction and the RS construction. In \fref{sec:gravity}, we review the computation of the graviton propagator in the five dimensional bulk, as well as a Gauss-Codazzi analysis, both concluding that gravity on the dark bubble is spin-2 and not spin-0. We then point out that uplifting gravitational waves on the dark bubble to 5d and computing their back-reaction on 4d results in the same metric as the backreaction considered purely in 4d, further showing consistency at perturbative order. In \fref{sec:holography}, we outline the holographic interpretation of the dark bubble, which clearly shows why localization is not an issue. In \fref{sec:FAQs}, we summarize the paper with a set of FAQs.

\section{Why the dark bubble is not an RS braneworld}\label{sec:differences}
Let us work in an effective theory of five dimensional Einstein gravity with matter
\begin{equation}
	S = \frac{1}{2κ₅²}\int \d x^5 \sqrt{-g^{(5)}} \left( R - 2 \Lambda + \mathcal{L}_m \right),
\end{equation}
where $κ₅² = 8πG₅$ is the reduced Newton's constant in 5d, and $\mathcal{L}_m$ is the matter Lagrangian. When $\mathcal{L}_m = 0$, the maximally symmetric spacetime solution is AdS₅, which can be written as
\begin{equation}
	\d s² = e^{2 A(z)}η_{μν} \d x^μ \d x^ν + \d z²\,,
\end{equation}
where the warp factor $A(z) \equiv \pm 2k z$, and $k \coloneqq \sqrt{-Λ/6}$ is the AdS curvature, and the holographic coordinate $z \in \left(-∞, ∞\right)$ increases monotonically from the center to the boundary of AdS.

Now consider $\mathcal{L}_m$ corresponding to a shell of matter located at $z=0$, with AdS₅ having curvature $k_\pm$ on either side of it i.e,
\begin{equation}
	A(z) = ϵ_+ k_+ Θ(z)z + ϵ_- k_- Θ(-z)z.
\end{equation}
This corresponds to a shell with constant tension
\begin{equation}
	\mathcal{L}_m = \frac{3}{κ₅²} \left(ϵ_+k_+ - ϵ_- k_- \right) δ(z)\,,
\end{equation}
which gives the tension of the shell. When $ϵ_- = - ϵ_+ = 1$, and $k_- = k_+ = k$, this corresponds to the RS scenario. In the absence of the $\mathbb{Z}₂$ symmetry that identifies $k_\pm$, and when the warp factor increases all the way from the center of AdS to the boundary i.e, $ϵ_+ = ϵ_- = 1$, we have the dark bubble. Since the radius of transverse spheres increases all the way from the center of the spacetime to the boundary, it is natural to think of the spacetime to the interior and the exterior of the bubble as an ``inside'' and an ``outside'' respectively. 

A naive dimensional reduction of this 5d action along the $z$ direction, gives the four dimensional Newton's constant. Integrating upto a cutoff surface $z=z_0$,
\begin{equation}
\begin{split}
	G₅ &= G₄ \int_{-∞}^{z_0} \d z e^{2 \left( ϵ_+ k_+ Θ(z) + ϵ_- k_- Θ(-z) \right)z}\\ 
	&= \frac{G₄}{2}\left(\frac{1}{ϵ_- k_-} - \frac{1}{ϵ_+ k_+}\right) + G₄ \frac{1}{2 ϵ_+ k_+} e^{2ϵ_+ k_+ z_0}\,.
\end{split}
\end{equation}
For RS, the second piece vanishes as $z_0 \rightarrow ∞$, and this gives the familiar result \cite{Randall:1999vf}
\begin{equation} \label{eq:g4rs}
	G_4^{\textrm{RS}} = G_5 k\,.
\end{equation}
For the dark bubble, on the other hand, the divergent second term is just the divergent volume of AdS as one approaches the boundary that be regulated, while the first term is regular as $z_0 →∞$, and gives the correct $G_4$. In the next section we will see how the Gauss-Codazzi equation, describing the backreacted embedding of the brane, automatically produces the correct, regulated value. 

As explained \cite{Banerjee:2020wov}, there is a negative sign that appears in the relation. The reason for this is that matter on the brane contributes with a negative sign to the effective energy density in 4D. However, back reaction from the bulk, contributing to the extrinsic curvature, yields a net positive energy density. We will come back to this below, when we discuss the Gauss-Codazzi equations. The net result is that the effective 4d Newton's constant is given by:
\begin{equation}\label{eq:g4dark}
	G_4^{\textrm{dark bubble}} = 2 G_5 \left(\frac{1}{k_+}-\frac{1}{k_-}\right)^{-1} =  G_5\frac{2 k_- k_+}{k_- - k_+} \,.
\end{equation}
It is crucial to realize that the dark bubble is not just the RS scenario without the $\mathbb{Z}_2$ symmetry. What is much more important is the difference in the choice of $ϵ_\pm$, as becomes apparent when one compares (\ref{eq:g4rs}) and (\ref{eq:g4dark}). In case of the dark bubble, you get a finite results only when $k_-$ differs from $k_+$. This is what you expect from a phase transition, with the nucleated brane separating two vacua with different cosmological constant. It is also the presence of the \emph{outside} AdS₅ that naively drives $G_4$ to zero. However, as explained in \cite{Banerjee:2020wov}, this is because unlike RS, our sources are strings extending in the fifth direction, which source non-normalizable gravitational modes in the exterior. Again, the best way to see this is through the Gauss-Codazzi equations.

\section{Gravity on the dark bubble}\label{sec:gravity}
\subsection{Local analysis using Gauss-Codazzi equations}
A purely gravitational way of understanding the four dimensional theory on the bubble wall is obtained by studying the induced Riemann curvature using the Gauss-Codazzi equations. This was done in \cite{Banerjee:2019fzz}. Below, we summarize the result of \cite{Banerjee:2019fzz} highlighting that it gives rise to 4d gravity. The Gauss equations are
\begin{equation}
	R^{(5)}_{αβγδ} e^α_a e^β_b e^γ_c e^δ_d = R^{(4)}_{abcd} + K_{ad}K_{bc} − K_{ac}K_{bd}\,,
\end{equation}
where Greek indices are in 5d, and Latin indices are in 4d. Using the thin-shell junction condition
\begin{equation}\label{eq:junction}
	8πG_5 S_{ab} = ΔK_{ab} − ΔK h_{ab}\,,
\end{equation}
and assuming that the extrinsic curvature on the brane is dominated by the five dimensional cosmological constant i.e, $K_{ab} = k h_{ab} + τ_{ab}$ with $τ_{ab}$ is sub-leading compared to $k$, this gives
\begin{equation}
\begin{split}
	&G^{(4)}_{ab} = h_{ab}\left[16πG_5σ\left(\frac{k_+k_-}{k_- - k_+}\right)-3k_+k_-\right]\\
	&+ \left(\frac{k_+k_-}{k_- - k_+}\right) \left[ \left(\frac{\mathcal{J}^{+}_{ab}}{k_+} - \frac{\mathcal{J}^{-}_{ab}}{k_-}\right) - \frac{1}{2}h_{ab}\left(\frac{\mathcal{J}^{+}}{k_+} - \frac{\mathcal{J}^{-}}{k_-}\right) \right]\,
\end{split}
\end{equation}
where $σ$ is the tension of the brane, and 
\begin{equation}
	\mathcal{J}_{ab}-R^{(4)}_{ab} = -3k²h_{ab}-k\left(2τ_{ab}+τh_{ab}\right) + \mathcal{O}\left(\frac{τ²}{k²}\right)\,.
\end{equation}
Let us now choose the bulk metric to be asymptotically AdS₅ with matter and a uniformly dense cloud of strings i.e.,
\begin{equation}\label{eq:AdS5metric}
	\d s_\pm² = -f(r)_\pm \d t² + \frac{\d r²}{f(r)_\pm} + r² \d \Omega_3²\,,
\end{equation}
with
\begin{equation}
	f(r)_\pm = 1 + k_\pm²r²- \frac{8G₅M_\pm}{3πr²} - \frac{2G₅\alpha_\pm}{r}\,.
\end{equation}
With this, the four dimensional Einstein tensor becomes
\begin{align*}
	&\left(G^{(4)}\right)^a_{b} =- \underbrace{2k_+ k_-\left(3 -\frac{8πG₅}{k_--k_+}σ\right)}_{\equiv 8πG_4\left(σ_\textrm{crit}-σ\right) \equiv  8πG_4Λ_4}
	δ^a_b \\
	&-\frac{8G₅}{πa(τ)⁴}\left(\frac{M_+k_- - M_-k_+}{k_--k_+}\right)
	\left(δ^a_0δ^0_b-\frac{1}{3}\sum_{i=1}^3δ^a_iδ^i_b\right)\\
	& -\frac{6G₅}{a(τ)³}\left(\frac{\alpha_+k_- - \alpha_-k_+}{k_--k_+}\right)δ^a_0δ^0_b\\
	&-16πG_5\left(\frac{k_+ k_-}{k_- - k_+}\right)\left(T_\textrm{brane}\right)^a_b \tag{\stepcounter{equation}\theequation}\\
    &= -8πG_4 Λ_4 \\ 
    &- \frac{4 G_4}{πa(τ)^4} \left( \frac{M_+}{k_+} - \frac{M_-}{k_-} \right) \left(δ^a_0δ^0_b-\frac{1}{3}\sum_{i=1}^3δ^a_iδ^i_b\right)\\
    & - \frac{3 G_4}{a(τ)^3} \left( \frac{α_+}{k_+} - \frac{α_-}{k_-} \right)δ^a_0δ^0_b
    - 8π G_4 \left(T_\textrm{brane}\right)^a_b\\ 
    &= 8π G_4 \left(T^{(4)} \right)^a_b\,.
\end{align*}
This is precisely the four dimensional Einstein equations, corresponding to a positive cosmological constant, radiation, matter and additional world volume matter on the brane respectively. In line with previous comments one notes that the last term $T_{\rm brane}$ contributes to the Einstein tensor with a {\it negative} sign to the energy density. In fact, it contributes with the same sign as the components on the \emph{inside} of the bubble, i.e., $M_-$ and $α_-$. Note that $S_{ab}$ in \eqref{eq:junction} is the total stress tensor on the shell, which includes $T_brane$ in addition to the matter, radiation and cosmological constant induced from the 5d bulk.

To understand why this is the physically correct sign, we start with the simple example of an AdS-Schwarzschild background with mass $M$. Assuming the same mass parameter $M$ on the two sides of the brane, we find a 4d cosmology with radiation, where the energy density is positive and proportional to $M/k_+ - M/k_->0$. Let us imagine that the black hole is replaced by a shell with the same mass $M$. The metric outside of the spherically symmetric shell will remain the same, as will the induced 4d physics. Let us now increase the radius of the shell so that it touches the brane. The matter can then be deposited onto the brane world with no change in the physics. Since the interior of the brane now is pure AdS, the term $-M/k_-$ is removed and replaced by the extra energy density on the brane, which contributes with the previously noted negative sign. This shows what is going on. It is a mistake to conclude that adding matter onto the brane gives rise to a negative energy density in 4d without any other consequences. What happens is that the added matter necessarily back reacts on the 5D space time, yielding a contribution to the extrinsic curvature that gives an overall {\it positive} 4d energy density.

Another amusing example involves the stretched strings pulling on the brane, which look like particles in 4D. If you put a mass on top of the brane, ignoring the back reaction discussed above, you would expect the brane to bend down towards the inside. In case of RS, due to the difference in sign, the bending is in the other direction, {\it away} from the inside as shown in \cite{Garriga:1999yh} (the same for both insides). The induced metric will be that of a positive mass. In RS you have two insides glued to each other, but for the dark bubble you have an inside and an outside. This means that the bending of the brane can be neatly explained by a pulling string. As shown already in \cite{Banerjee:2018qey}, if you have a string cloud, corresponding to dust in 4D, the pulling is fully accounted for by the extrinsic curvature, reproducing FRW with a positive mass density. 

\subsection{Graviton propagator in momentum space}
To understand the nature of gravity on the dark bubble, one can compute the graviton propagator between two point particles on the 4d bubble, just like in \cite{Giddings:2000mu}. This was done in \cite{Banerjee:2020wix}, and confirms the presence of Einstein (spin-2) gravity on the dark bubble. Let us summarize the result of that computation below.

Consider perturbations $h_{ab}$ of the bulk metric in Poincarè coordinates
\begin{equation}
	\d s² = \d ξ² + a(ξ)² \left(η_{ab} + h_{ab}\left(ξ,x^a\right)\right)\d x^a \d x^b\,,
\end{equation}
where $a(ξ) = \exp(k ξ)$, with $k$ being the AdS₅ curvature.
Choosing $h_{ab}$ to be in the Lorenz gauge $∂^ah_{ab} = \frac{1}{2}∂_b h$, and defining $\overline{γ}_{ab} = γ_{ab} - \frac{1}{2}η_{ab} γ$, where $γ_{ab} \coloneqq a(ξ)²h_{ab}$ gives, after a Fourier transformation
\begin{equation}
	\left( -\frac{p²}{a²} + ∂_ξ² - 4k²\right) \tilde{χ}_{ab}(p,ξ) = -2κ₅²\tilde{\Sigma}_{ab}\,,
\end{equation}
where $χ_{ab},\Sigma_{ab}$ are the traceless pieces of $\overline{γ}_{ab}$ and the stress tensor $T_{ab}$, and tildes are their Fourier transformations in the transverse directions. We can solve for the corresponding Green's function $Δ_{\tilde{χ}}\left(p;a_+, a_-\right)$ (where $a_\pm$ are the scale factor inside and outside the shell) piece-wise, inside and outside the shell in terms of modified Bessel functions $K_2$ and $I_2$. These diverge at large and small $a$ respectively. Regularity at the Poincarè horizon ($a→0$) excludes $I_2$ inside the shell, while the presence of string sources stretching out in the fifth direction allows for both $K_2$ and $I_2$ outside the shell.
{\small
\begin{equation}\label{eq:choice_bessel}
	\begin{split}
		\Delta_{\tilde{χ}}^+(p;a_+,a_-) &= A(p,a_-) K_2\left(\frac{p}{a_+ k_+}\right)+B(p,a_-) I_2\left(\frac{p}{a_+ k_+}\right),\\
		\Delta_{\tilde{χ}}^-(p;a_+,a_-) &= C(p,a_+) K_2\left(\frac{p}{a_- k_-}\right)\,.
	\end{split}
\end{equation}
}
When the momentum is small compared to the radius of the bubble and the curvature, i.e. $p \ll ak$, the Bessel functions can be expanded as
\begin{equation}
\begin{split}
   &\lim\limits_{a→∞} K_2\left(\frac{p}{a k}\right) =\frac{2 a^2 k^2}{p^2}-\frac{1}{2}\\
   &\quad +\frac{4 p^2 \log (a)-4 p^2 \log \left(\frac{p}{2k}\right)-4 \gamma  p^2+3 p^2}{32 a^2 k^2}+\ldots,\\
    &\lim\limits_{a→∞} I_2\left(\frac{p}{a k}\right) = \frac{1}{8}\frac{p²}{a²k²}+\ldots\,,
\end{split}
\end{equation}
where $\gamma$ is the Euler-Mascheroni constant.
The thin-shell junction conditions then determine $A$, and $C$. Expanding the Bessel functions close to the boundary, $a→∞$, and requiring that the massless scalar modes of AdS₅ vanish, further fixes $B$. In the small momentum limit, this becomes
\begin{equation}\label{eq:K2-outside}
 \Delta_{\tilde{χ}}^{\rm{shell}}(p; a_s,a_s) = \frac{a_s²}{p²}\left(\frac{2k_- k_+}{k_- - k_+}\right)+\mathcal{O}\left(p^0\right).
\end{equation}
The leading order term gives the zero mode of the five dimensional graviton with sub-leading corrections in finite momentum. This 5d zero mode corresponds to the four dimensional graviton, confirming Einstein's gravity on the dark bubble.
Fourier transforming this back into position space gives
\begin{equation}
    Δ_5 =\int \frac{\d^4 p}{(2π)^4} e^{i p \left( x- \tilde{x} \right)} \frac{Δ_{\tilde{χ}}}{a^2} 
    = \frac{κ_4²}{κ_5²}\frac{1}{r} + \ldots\,,
\end{equation}
where $r$ is the radial coordinate in 4d.
Following \cite{Giddings:2000mu}, $χ_{ab}$ is the convolution of the scalar Green's function with the source,
\begin{equation}
\label{eq:chiab}
    χ_{ab} = - 2κ_4² \int \sqrt{g}\, \Sigma_{ab} Δ_5\,.
\end{equation}
So far we have assumed that the brane sits at $ξ=0$. However, as discussed in \cite{Banerjee:2020wix}, the brane bends in response to a source; the result is that placing matter directly on the brane (or having a string pulling from inside) contributes with a negative sign to the stress tensor. 
As a result, the brane sits at $ξ=f(x^a)$ instead of $ξ=0$. Performing a coordinate transformation $ξ↦ξ-f(x^a)$, and a corresponding change in the world volume coordinates $x^a$, brings back the brane to $ξ=0$ in these new coordinates.
This contributes additional terms (proportional to the bending) to the metric perturbation $χ_{ab}$, as well as to the stress tensor $\Sigma_{ab}$. 
These extra contributions are not traceless.
Demanding that the  full stress tensor, including the contribution from bending, be traceless then determines the amount of bending. 
All of this was taken into account and the metric perturbation was computed in momentum space in \cite{Banerjee:2020wix}. The computation can also be performed in the position space, and leads to the analog of \cite[eq (3.25)]{Banerjee:2020wix}
\begin{equation}
    h_{ab} = - 2κ_4² \int \sqrt{g} \left( T_{ab} - \frac{1}{2} η_{ab} T \right) \frac{1}{r} + \ldots \,.
\end{equation}
With the negative sign as discussed above, this correctly reproduces 4d Einstein gravity.

The result of the above computation crucially depends on the choice of boundary conditions. 
If one does not allow for the growing modes i.e, $A=0$ in \eqref{eq:choice_bessel}, then \eqref{eq:K2-outside} becomes,
\begin{equation}
\label{eq:noK2outside}
\begin{split}
    &Δ_{\tilde{ξ}}^{\rm shell} = \left[ \frac{K_1\left(\frac{p}{a_s k_-}\right)}{K_2\left(\frac{p}{a_s k_-}\right)} + \frac{I_1\left(\frac{p}{a_s k_+}\right)}{I_2\left(\frac{p}{a_s k_+}\right)}\right]^{-1}\\
    &\Rightarrow  \lim_{p \ll a k} Δ_{\tilde{ξ}}^{\rm shell} =
    \frac{1}{4k_+} + \mathcal{O}\left(p²/a²\right)\,,
\end{split}
\end{equation}
which is exactly the result obtained in \cite[eq (2.28)]{Mirbabayi:2022eqn}, and leads to gravity mediated by a scalar field, instead of actual Einstein gravity. The precise difference between the two lies in the fact that, in case of the scalar mediated gravity, while the scalar modes yield a leading behaviour similar to \eqref{eq:K2-outside}, the tensor modes limit to a constant as in \eqref{eq:noK2outside}. However, as we demonstrated above, turning on a non-vanishing $A$, i.e. including $K_2$ on the outside, reproduces the desired $1/p²$ behavior of the propagator for the tensor modes as well and therefore can be interpreted as $4$-dimensional Einstein gravity on the shell, beyond any ambiguity. 

This emphasises, once again, how important it is to incorporate strings stretching out in the fifth direction in the bulk, to realise real Einstein's gravity on the $4$-dimensional bubble wall.
These string sources make our model fundamentally different from RS, as well as a naive KK reduction along the 5th direction, where the sources are point particles in four dimensions, with their wave functions smeared along the fifth dimension. While this introduces an apparently unwelcoming non-localization of gravity, the scale of violation of the localization is controlled by the AdS length scale, and can be microscopic. Therefore, in principle, one can put a cutoff at the AdS throat and be content with the effective theory of Einstein's gravity on the bubble wall coupled with matter generated by the endpoints of the strings. The mass of the matter fields turns out to be independent of the size of the extra dimension through mass renormalization \cite{Banerjee:2020wov} as the bubble traverses along the extra dimension. 

The novel advantage of this tiny sacrifice is four-fold. {\it First}, as we just demonstrated above, this is the most natural way to realize an effective Einstein's gravity on a dS bubble wall. {\it Second}, this allows for the possibility to model an effective theory of gravity coupled to the standard model particles \cite{Banerjee:2019fzz}, while for the latter one does not need to introduce any extra brane as in the two-brane RS model \cite{Randall:1999ee}. Furthermore, as we showed in \cite{Banerjee:2020wov}, the residual time dependence of the matter coupling due to the time-dependent radion field \cite{Montefalcone:2020vlu, Karch:2020iit} is automatically taken care of by the dynamics of the bubble. {\it Third}, the dynamics of the our bubble universe turns out to be perfectly consistent with Vilenkin's tunneling proposal in quantum cosmology, with the bubble nucleation in $5$ dimensional AdS spacetime being identified with a $4$ dimensional ``Big Bang'' event. This was established in 
\cite{Danielsson:2021tyb} by showing that the amplitude of bubble nucleation in $5$ dimensions matched identically with Vilenkin’s tunneling amplitude in $4$ dimensional quantum cosmology. {\it Last but not the least,} the aforementioned tunneling dynamics was nicely embedded in a stringy model in \cite{Danielsson:2022lsl}, where an unstable AdS configuration with large enough R-charge chemical potential, emits color D3 branes and tunnel through a potential barrier to a low energy stable configuration. At lowest order, the universe residing on the worldvolume of the emitted D3 brane has a vanishing cosmological constant, and keeps growing up to a maximum size before it starts contracting again. Taking $1/N$ corrections into account, consistent with the WGC, one obtains a small cosmological constant compatible with observations.

\subsection{5d uplift of 4d gravitational waves}
In the presence of spin-2 gravity on the four dimensional dark bubble, one would expect to find tensor gravitational waves. These solutions were explicitly found in \cite{Danielsson:2022fhd}, where they further constructed the five dimensional gravitational waves that these uplift to. Let us summarize their results below.

In a transverse-traceless gauge, four dimensional gravitational waves on the dark bubble (conformally $ℝ^1 × S^3$) are given by $h_{ij} = h^{\rm 4d}(η) Y_{ij}$, where $η$ is conformal time on the dark bubble, and $Y_{ij}$ are tensor spherical harmonics on $S^3$. $h(η)$ is a linear combination of
\begin{align*}
    h_1^{\rm 4d}(η) &= \frac{\cos{((n+1)η)}}{n+1} + \sin{η} \sin{(nη)}\,,\\
    h_2^{\rm 4d}(η) &= \frac{\sin{((n+1)η)}}{n+1} - \sin{η} \cos{(nη)}\,,
\end{align*}
where $3-n²$ is the eigenvalue of the Laplacian on $S^3$. The gravitational waves, which perturb the metric at the first order, source an energy-momentum tensor. At late times, this corresponds to curvature and radiation.

Gravitational waves in AdS$_5$ which are the uplift of the four dimensional waves are of the form $h_{ij} = h^{\rm 5d}(t, z)Y_{ij}$, where $t$ and $z$ are the global time coordinate, and the radial direction in AdS$_5$ respectively (like in \eqref{eq:AdS5metric}, but with $M_\pm=α_\pm=0$). $h^{\rm 5d}$ is a linear combination of
\begin{widetext}
\begin{align*}
    &h^{\rm 5d}_1(t,z) = \frac{(k z)^{n-1}}{\left(1+k^2z^2\right)^{\frac{n-1}{2}}}
    \left[\frac{\frac{1}{2}(1+n)(2-n)+k^2z^2}{(n+1)\left(1+k^2z^2\right)}\cos\left((n+1)kt\right)+\sin(kt)\sin(nkt)\right]\,,\\
	&h^{\rm 5d}_2(t,z) = \frac{(k z)^{n-1}}{\left(1+k^2z^2\right)^{\frac{n-1}{2}}}
    \left[\frac{\frac{1}{2}(1+n)(2-n)+k^2z^2}{(n+1)\left(1+k^2z^2\right)}\sin\left((n+1)kt\right)-\sin(kt)\cos(nkt)\right]\,.
\end{align*}
\end{widetext}
Similar to 4d, this perturbation induces an energy momentum tensor in AdS$_5$ and has components corresponding to curvature, radiation and flux. This sources a back reaction on the 5d metric, which induces a second order correction to the stress tensor on the bubble. \cite{Danielsson:2022fhd} showed that the second order correction induced on the dark bubble from this five dimensional back reaction is exactly the same as what would have been obtained by directly solving Einstein's equations in four dimensions directly at second order. This is a non-trivial result that further establishes the consistency of the dark bubble model pertubatively in response to gravitational wave fluctuations.

\section{The view from holography}\label{sec:holography}

 Not appreciating how the dark bubble differs from RS, and applying the intuition gained from RS, invariably leads astray when trying to understand the physics of the model. It is true that the dark bubble as well as RS imply Einstein spin-2 gravity in the low energy limit, and that too for similar technical reasons. But the role of the fifth dimension, and the way matter is introduced, is completely different. In RS, matter as well as gravity is localized to the brane, and even observers can be viewed as stuck on the brane. Furthermore, the two sides of the brane are best viewed as two \emph{insides}. In case of the dark bubble, sources are not localized to the brane but extend outwards in the bulk all the way to infinity. Keeping the philosophy of RS in mind this seems confusing. However, the correct way to think of the dark bubble is holographically from the point of view of AdS/CFT. 
 
 It is well known, see e.g. \cite{Maldacena:1998im,Rey:1998ik}, that point sources like quarks can be represented as hanging strings. There is no issue of localization in the fifth dimension, since the extra dimension corresponds to scale. The connection can be made even more precise by imagining that we start in holographic AdS before the dark bubble has nucleated. We then have holography of the usual kind with no 4d gravity, albeit in an unstable background. When the brane nucleates, not only is the universe created in the sense of quantum cosmology, we also get the force of gravity induced by the presence of the brane. Note that from the point of view of holography, gravity is only present if you go to large enough scales, while in the extreme UV, there is no gravity.

Even though a string stretching outward from the brane along the fifth direction can be infinitely long, the effective 4d mass of its end point on the brane is proportional to the product of the 5d AdS length and the tension of the string, and is therefore finite \cite{Banerjee:2020wix}. In fact, in the presence of a cloud of strings in 5d, the induced metric on the dark bubble is that of Friedmann cosmology with dust. Furthermore, the force between the strings is such that the geodesic motion in 5d is compatible with the 4d geodesic motion of its end points. While all of this can be done with semi-infinite strings, it is also possible to regulate the length of the string (and thus the volume of external AdS$_5$) in a natural way by introducing a cutoff in the form of an RS brane in the UV. 
While the graviton zero mode grows up and away from the dark bubble (threatening to be non-normalizable), it decays down and away the RS brane (hence normalizable). The graviton zero mode therefore now has a finite norm. The question instead is: where is 4d gravity? On the RS brane or on the dark bubble?

\begin{figure}
    \centering
    \def\svgwidth{1.25\linewidth}
    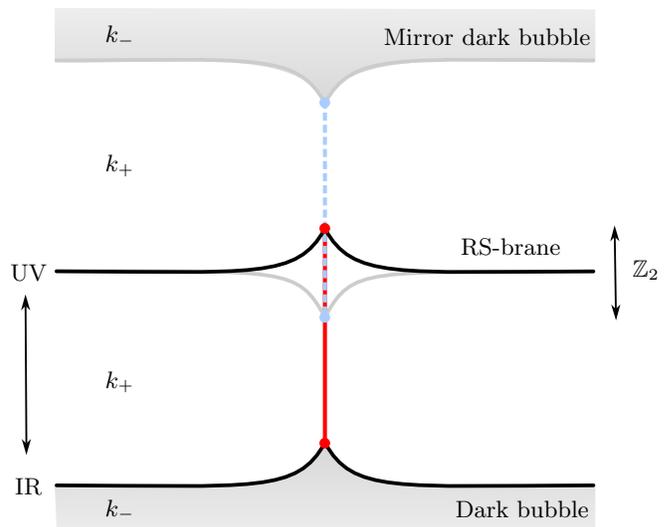
    \caption{The dark bubble brane in the IR and an RS brane in the UV to normalize the graviton zero mode. Owing to the $\mathbb{Z}_2$ symmetry across the RS brane, there is a second copy of the dark bubble on top (labelled ``mirror dark bubble'' and shown in gray). A string stretching between the branes along the fifth direction (shown in red) causes the branes to bend. The mirrored string is shown with a dotted blue line, and the corresponding bending is shown in gray. In the physically relevant situation discussed in the explicit realization of \cite{Danielsson:2022lsl}, $G_4^{\textrm{dark bubble}} \gg G_4^{\textrm{RS}}$, and 4d gravity is realized on the dark bubble with the RS brane serving as a spectator.}
    \label{fig:2brane}
\end{figure}

Let us consider a situation in \fref{fig:2brane} in a little more detail, which has strings stretching between the dark bubble brane in the IR and the RS brane in the UV. The strings pull on the IR brane, where they locally look like particles of finite mass independent of the length of the strings, and move as if affected by $G_4^{\textrm{dark bubble}}$. Bending of dark bubble sources growing (formerly non-normalizable) modes that increase all the way up to the cutoff UV brane, where they induce a bending (or gravitational field) that is enhanced by the relative scale factor between the two branes (See \cite{Banerjee:2020wov} for details). This makes sure that the strings, independent of their lengths, move in sync with their endpoints on the dark bubble. The distance between the end points is scaled but so is the strength of gravity. In fact, on the UV brane it looks like there is a lot more mass around since gravity has the same strength at a larger distance. This is just the mass of the full string. This mass is not directly on the brane but is sensed by the UV brane precisely because of the cutoff non-normalizable modes. The independence of the length of the string from the point of view of the IR-brane has a parallel also in flat space Kaluza-Klein reduction. In a compactification of a circle of radius $R$ with wound strings as sources, one has $G_4= G_5/R$ and $M=τ R$. So in $G_4 M =G_5 τ$, (the Schwarzschild radius) the length of the compact circle $R$ drops out. All of this can also be understood in parallel with a standard 2-brane (Minkowski) RS setup \cite{Randall:1999vf} with a UV brane and a (negative tension) IR-brane by setting $k_- → ∞ $ in \eqref{eq:G4s}. In this special case, everything follows through with $G_4^{\textrm{\scshape rs}} = G_4^{\textrm{dark bubble}}$. \footnote{We thank Andreas Karch for extensive discussions about this.}

For the dark bubble there is a further important twist. There is no reason why the 4d Newton's constant on the RS brane and the dark bubble brane should be the same. In fact,
\begin{equation}\label{eq:G4s}
    G_4^{\textrm{RS}} =k_+ G_5,\qquad
    G_4^{\textrm{dark bubble}} = \frac{2k_- k_+}{k_- - k_+}G_5\,.
\end{equation}
Let us use the explicit embedding proposed in \cite{Danielsson:2022lsl}. With the physically motivated limit $k_- \sim k_+ \sim k$, with $k_- -k_+ \sim k/N_c$, one gets
\begin{equation}
    G_4^{\textrm{RS}} = \frac{1}{N_c} G_4^{\textrm{dark bubble}}\,.
\end{equation}
With the suggested value $N_c \sim 10^{60}$, this gives $G_4^{\textrm{dark bubble}} \gg G_4^{\textrm{RS}}$. This means that any matter localized on the RS-brane will have a completely subleading contribution to the curvature of the brane. Everything is completely dominated by the stretched strings that contribute with their full length and a Newton's constant that is much larger. It is therefore $G_4^{\textrm{dark bubble}}$ that should be identified with the physical Newton's constant in 4d. The role of the RS brane is reduced to just a spectator, and serves little more than to regulate the norm of the graviton zero mode. Another peculiar and very interesting feature of the dark bubble construction is the \emph{inverted hierarchy} of Planck scales:
\begin{equation}
    \ell_4^{\textrm{dark bubble}} \gg \ell_5 \gg \ell_4^{\textrm{RS}}\,,
\end{equation}
as compared to KK or RS compactification which both yield $\ell_5 > \ell_4$. For more details on this and how a small cosmological constant arises in the dark bubble model, see \cite{Danielsson:2022lsl}. 

In summary, the dark bubble model can either be viewed as RS with an extra brane in the IR, or as a dark bubble in the IR with a cutoff RS brane in the UV---both of which are perfectly finite (see \fref{fig:2brane}). 
How does one study the physics of the setup? 
One option would be to focus on the UV RS brane. 
One can add matter there, identify its induced metric with 4d space-time, and see of what possible use, if any, the IR brane would be.  This is the conventional viewpoint. For this to make sense, one must explicitly forbid, e.g., strings stretching towards the IR brane that otherwise would dominate the physics.  
Another option would be to focus on the dark bubble brane in the IR, identify its induced metric with 4d space-time, and add matter using structures extending between the branes in 5d. As we have seen, the influence of the RS brane (beyond serving as a cutoff) is small---at least at low energy.

Coming back to the dark bubble, what happens when it expands? If the strings are far enough not to gravitate, their endpoints on the dark bubble will move away from each other as the brane expands and eats away on the strings.
On the UV brane, there is no change in the distance between the endpoints (if they are far enough not to be gravitationally bound). Instead, from the perspective of a 4d observer on the UV brane, the masses will decrease (and so will the spatial size of any bound object). This is an alternative, conformally transformed, description of the same 4d universe as described by the dark bubble observer, where the galaxies, instead of moving away from each other, are getting smaller. 
On the other hand, the observer on the dark bubble finds a parametrization that looks exactly like matter in an expanding dS$_4$ universe. 

We believe that focusing on the IR brane, with sources holographically extending from the IR to the UV, is true to holography and in the spirit of renormalization. It is a matter of choosing the right observables and initial conditions---to this end, we set up the observables in the IR and then arrange the UV to match it. We find the results from this approach encouraging.

\section{FAQs} \label{sec:FAQs}
Let us summarize, by addressing some of the most common confusion regarding the dark bubble model that we have encountered over the years.
\begin{itemize}[leftmargin=*]
    \item \textbf{Is the dark bubble just a mismatched RS setup?}\\ 
    No. As discussed in \cite{Banerjee:2018qey,Banerjee:2019fzz,Banerjee:2020wix,Banerjee:2020wov}, and highlighted in \fref{sec:differences}, apart from having different cosmological constants on either side of the brane, a crucial difference is that the dark bubble has an \emph{inside} and an \emph{outside} in contrast to the two insides of RS. This changes the way gravity is realized on the dark bubble, and necessitates the presence of a growing mode in the bulk spacetime outside the bubble. This, then, leads to 4d matter being endpoints of strings stretching along the 5th dimension, in contrast to RS where 4d matter is localized on the brane. The response of the brane to the presence of matter is similar in both models, but the presence of stretched strings makes all the difference as emphasized above.
    \item \textbf{Is gravity on the dark bubble  spin-2 or spin-0?}\\
    Gravity on the dark bubble is spin-2. The confusion around this stems again from wrongly assuming that gravity is localized, and not considering that sources, such as the stretched strings, holographically extend in the fifth direction. In their absence, the graviton propagator only has a mode that decays at the boundary of the outside AdS$_5$ and the graviton indeed appears to be spin-0 as recently obtained in \cite{Mirbabayi:2022eqn}. However, the dark bubble necessarily needs extended sources to produce 4d gravity. These source growing modes in the bulk outside of the bubble, which then gives a spin-2 graviton. This is further confirmed by a Gauss-Codazzi analysis, as well as the consistent 5d uplift of 4d gravitational waves and its backreaction in 4d as discussed above. Therefore, there is no doubt that gravity on the dark bubble in the presence of stretched strings is indeed spin-2.
    \item \textbf{Shouldn't gravity be localized to the brane as is the case for RS?}\\
    Gravity is {\it not} localized, as argued above it cannot be. The correct way to think of the extra dimension, transverse to the dark bubble, is through holography. Sources extend all the way to the UV at infinity, and affect the space time all the way through non-normalizable modes. As explained in section 4, it is possible to introduce a cutoff to restore normalizability. The low energy physics is not affected by this.\item \textbf{Is the sign of the energy on the brane wrong?}\\
    The sign is correct. Just as you cannot change the mass of the Earth, without affecting the space time metric, you cannot add matter to the brane without the metric changing. The full back reacted solution needs to be considered, and this leads to a net {\it positive} energy density, when the extrinsic curvature is taken into account through the junction condition.
    \item \textbf{There is no convincing embedding of the dark bubble into string theory.}\\
    This is wrong. \cite{Danielsson:2022lsl} describes an explicit embedding, with a positive cosmological constant that is naturally small. In RS, the tension of the RS brane needs to be {\it increased} beyond its critical value through an uplift. This is essentially what is going on in KKLT, which is criticized within the swampland program. In case of the dark bubble it is completely different. As soon as you have a tension that is {\it smaller} than the critical value you automatically get dS.\footnote{Interestingly, the dark bubble model does not have much to say about the swampland distance conjecture\cite{Ooguri:2006in} or the trans-Planckian censorship conjecture \cite{Bedroya:2019snp}. Moduli corresponding to the five internal dimensions beyond the AdS$_5$, should be stabilized, while the radius of the brane bubble expanding in the fifth dimension corresponds to the scale factor in the 4d cosmology. The bubble wall moves a microscopic distance (order AdS length) in the fifth dimension, over large e-folds of expansion in 4d. Since the distance traversed never exceeds a few AdS lengths, questions of runaway moduli or approaching the boundaries of moduli space never arise.}
    
    \item\textbf{Does the dark bubble decay instantaneously? How long before it collides with another bubble?}\\
    No, an observer living on the dark bubble sees that it lasts longer than a Hubble time. The time scale relevant for the observers is the proper time of the brane. For a large bubble, there is a huge blue shift due to the metric factor, making sure that cosmological times will pass on the brane before the bubble hits another bubble. At least if the probability of nucleation is sufficiently small. In fact, as shown in \cite{Banerjee:2019fzz}, a Hubble time will pass while the proper radius of the dark bubble only increases by the AdS-radius. One should note however, that this crucially depends on the choice of boundary conditions. Assuming that the bubble exists at current time, it enjoys a long life time. This cutoff in time is equivalent to choosing a spatial cutoff where the boundary of AdS is excised away by the presence of an RS brane close to it.

\end{itemize}

\section{Acknowledgments}
We are extremely grateful to Andreas Karch for an absolutely fascinating and incredibly useful discussion about several aspects of the dark bubble model. We would also like to thank Mehrdad Mirbabayi for correspondence; and Rob Tielemans, Alessandro Tomasiello, and Thomas Van Riet for useful discussion. SG is supported in part by INFN and by MIUR-PRIN contract 2017CC72MK003.

\bibliography{references}
\bibliographystyle{apsrev4-2.bst}

\end{document}